\documentclass[conference]{IEEEtran}
\IEEEoverridecommandlockouts
\usepackage{verbatim}
\usepackage[utf8x]{inputenc}
\usepackage[pdftex]{graphicx}
\usepackage{epstopdf}
\usepackage{tabularx}
\usepackage{booktabs}
\usepackage[table]{xcolor}
\usepackage{booktabs}
\usepackage{ctable}
\usepackage{multirow}
\usepackage[cmex10]{amsmath}
\usepackage{algorithm}
\usepackage{caption}
\usepackage{subcaption}
\usepackage{algpseudocode}
\usepackage{pifont}
\usepackage{amssymb}
\usepackage{amsthm}
\usepackage{cite}
\usepackage{textcomp}
\usepackage{array}
\usepackage{enumerate}

\newtheorem{theorem}{Theorem}
\newtheorem{corollary}{Corollary}

\newcommand{\B}{\operatorname{B}}
\newcommand{\IoT}{\operatorname{IoT}}
\newcommand{\I}{\operatorname{I}}
\newcommand{\PN}{\operatorname{PN}}
\newcommand{\R}{\operatorname{R}}

\begin{document}
\bstctlcite{IEEEexample:BSTcontrol}
\title{Spectrum Sharing Protocols based on Ultra-Narrowband Communications for Unlicensed Massive IoT} 
\author{\IEEEauthorblockN{Ghaith Hattab and  Danijela Cabric}
	\IEEEauthorblockA{Department of Electrical and Computer Engineering, University of California, Los Angeles\\
Email: ghattab@ucla.edu, danijela@ee.ucla.edu}
\thanks{This work has been supported by the National Science Foundation under grants 1527026 and 1149981.}
}

\maketitle


\begin{abstract}
Ultra-narrowband (UNB) communications is an emerging paradigm that tackles two challenges to realizing massive Internet-of-things (IoT) connectivity over the unlicensed spectrum: the intra-network sharing, i.e., how the spectrum is shared among IoT devices, and the inter-network sharing, i.e., the coexistence of the IoT network with other incumbent networks. Specifically, intra-network sharing is enabled by using extremely narrowband signals to connect a massive number of IoT devices without any prior network synchronization. Further, to enhance robustness to incumbent interference, each IoT packet is sent multiple times, each at a different frequency within a single band. Nevertheless, the interplay between intra-network sharing and inter-technology coexistence at a large scale remains unclear. Thus, in this paper, we develop an analytical framework to model and analyze UNB networks. We use stochastic geometry to derive the probability of successful transmission, identifying the impact of intra- and inter-network interference on the performance. In addition to analyzing the existing single-band access protocols, we present two multiband schemes, where each BS listens to a single band for practical implementation. Different access protocols are further compared in terms of the transmission capacity, i.e., the maximum number of IoT devices a UNB protocol can support in the presence of incumbent networks. Simulation results are provided to validate the derived closed-form expressions. It is shown that the diversity achieved by sending multiple transmissions is beneficial when the interference is dominated by the incumbent network. Further, multiband access, where each packet is sent over a different band, provides an additional diversity gain, enabling the UNB network to support a very large number of IoT devices. The gains, in success probability and transmission capacity, are higher when devices are not restricted to connect to a single base station.
\end{abstract}


\begin{IEEEkeywords} 
Internet of Things, LPWA, massive IoT, spectrum sharing, stochastic geometry, success probability, transmission capacity, ultra-narrowband.
\end{IEEEkeywords}


\section{Introduction}\label{sec:introduction}
The massive Internet-of-things (IoT) market segment opens a myriad of opportunities and applications that help build smarter cities via utility metering, traffic management, infrastructure monitoring, etc \cite{DawyYaacoub2017,Zanella2014}. Such applications are enabled by deploying a massive number of low-cost battery-powered IoT devices, e.g., sensors, machines, and automated devices, at a large scale. In addition, these IoT devices sporadically transmit few small packets per day, mitigating the need for high-throughput links \cite{DawyYaacoub2017}. For these reasons, a new class of networks, known as low-power wide-area (LPWA) networks, has begun to emerge in recent years to meet the unique requirements of massive IoT \cite{Raza2017,CentenaroZorzi2016,Xiong2015}. LPWA networks rely on the use of the unlicensed spectrum due to its low capital expenditure, provide long-range connectivity, primarily using bands sub-1GHz due to their favorable propagation conditions, and use lightweight access protocols to limit the communication overhead and extend the lifetime of end-devices \cite{Raza2017}.

To realize massive IoT connectivity using LPWA networks over the unlicensed spectrum, there are two key challenges that must be addressed. The first one is the \emph{intra-network sharing}, i.e., how the spectrum is shared among IoT devices belonging to the same network, particularly in the absence of synchronization between IoT devices and the network. The second challenge is the \emph{inter-network sharing}, as the unlicensed spectrum is typically occupied by other incumbent networks, with more powerful end-devices compared to the low-cost IoT devices. These sharing challenges have motivated the emergence of the ultra-narrowband (UNB) network, a variant of LPWA networks \cite{Chen2017}.

\begin{table}[!t]
	\caption{Examples of UNB technologies}
	\label{tab:UNB}
	\centering
	\small 
	\begin{tabular}{|l|c|}
		\hline
		\textbf{UNB Technology }  				&  \textbf{Bandwidth (Hz)}\\\hline
		\multirow{2}{*}{Sigfox \cite{Sigfox2017}  } & 600 (US)\\	
		& 100 (Europe)\\	\hline
		WavIoT NB-Fi \cite{WAVIoT2016}   & 100\\\hline
		NWave Weightless-N \cite{Weightless2018} &200\\\hline
		Telensa \cite{Martindell2013} &500\\\hline 
	\end{tabular}
\vspace{-0.1in}
\end{table}

In UNB networks, data communication is done using extremely narrowband signals to connect a large number of devices and improve interference robustness by concentrating the signal energy into ultra-narrow channels. Indeed, the uplink (UL) signal bandwidth is typically in the range of 100-600Hz as shown in Table \ref{tab:UNB}. In addition, these networks rely on simple ALOHA-like access protocols, where IoT devices avoid associating and synchronizing  with any UNB base station (BS), i.e., devices broadcast their packets at any time and frequency. The only restrictions on these devices are primarily related to how many packets per day each can send and the bandwidth of the band that comprises the different channels a device can pick from. Furthermore, to combat the absence of acknowledgments from the network and the presence of incumbent networks, signal repetition is used, where each IoT packet is sent multiple times, one after another, each at a different frequency within the predetermined multiplexing band. The packet is successfully transmitted if any BS decodes any of the packet transmissions, yet no combining of these packets is done nor cooperation with other BSs is used. In this paper, our objective is to model and analyze UNB access to understand its effectiveness in terms of intra-network sharing and its resilience to interference from incumbent networks. 

Existing works on modeling and analyzing UNB networks have focused on simple network set-ups or have ignored some features of UNB networks. For instance, in \cite{Mo2016a,Do2014,Li2017}, the network is assumed to have a single BS and each message is sent once. In \cite{Mo2016}, a single BS is considered and the impact of sending the same packet multiple times is studied in terms of the probability of collision, i.e., two or more devices picking the same time-frequency resources. Since these works do not study large-scale networks, the intra-sharing capabilities of UNB networks remain unclear, particularly because a transmission of an IoT device can still be successful even if  another device has sent its packet over the same time-frequency slot. In addition, no prior work has analyzed the UNB network when it shares the spectrum with other networks. For these reasons, a more comprehensive analytical framework is needed.

The main contributions of this paper are twofold. First, we present an analytical framework to model UNB networks in the presence of other interfering networks using stochastic geometry \cite{Haenggi2012}, which has become a powerful tool to model  cellular networks \cite{ElSawyWin2017,HattabCabric2016a}. Since UNB networks rely on devices transmitting at random times and frequencies, stochastic geometry can help capture the spatial randomness of these networks and obtain theoretical expressions of key performance indicators. In particular, we derive the success probability in closed form, which is defined as the probability of having at least one of the messages successfully decoded. Different from prior work that uses collision as a performance metric, success probability captures the spatial variations of the signal-to-noise-plus-interference ratio (SINR). In addition, it helps determine the transmission capacity \cite{Hunter2008,Weber2012}, defined as the maximum density of IoT devices that can be supported in the network for a given success probability constraint. Second, existing access protocols assume devices send signals over narrowband channels within a single multiplexing band. Thus, one key question is whether using multiple bands can affect the intra-network sharing capabilities of the UNB network, given that each BS is restricted to listen to just one of the bands. In this paper, we present two different multiband spectrum sharing access protocols, analyze them, and compare them with the single-band access scheme and a benchmark protocol, where BSs can listen to all multiplexing bands. Further, for all access protocols, we consider two different signal repetition schemes: random and pseudorandom (PN), and two different device-BS association scenarios: nearest BS association and no BS association. Several design guidelines are gleaned from the analysis and further validated via Monte Carlo simulations. 

The rest of the paper is organized as follows. We present the system model, the access protocols, and the key performance metrics in Section \ref{sec:model}. Performance analysis of all protocols is given in Section \ref{sec:analysis}. Simulation results are presented in Section \ref{sec:simulations} and the conclusions are drawn in Section \ref{sec:conclusion}.


\section{System Model and Access Protocols }\label{sec:model}

\subsection{Network topology}
We consider a random spatial topology of BSs, IoT devices, and interfering networks. Specifically, for the UNB network, we assume that BSs' locations are generated from a homogeneous Poisson Point process (HPPP) $\Phi_{\B}$, with density $\lambda_{\B}$, whereas IoT devices are generated from another independent HPPP $\Phi_{\IoT}$ with density $\lambda_{\IoT}$. IoT signals are transmitted at power $P_{\IoT}$, occupying a bandwidth $b$. Each signal is repeatedly sent $N$ times, one after another. The temporal generation of IoT traffic is modeled as $\lambda_{\operatorname{T}}=t/T$, where $t$ is the duration of a single message and $T$ is the period between two messages. In addition, we consider an incumbent interfering network, where the locations of transmitters are generated from an independent HPPP $\Phi_{\I}$ with density $\lambda_{\I}$, and each interferer transmits at power $P_{\I}$ over a bandwidth $B_{\I}\gg b$, where this bandwidth is assumed to be overlapped with the spectrum used by the UNB network. The interfering network could be a WiFi network or another IoT-based network, e.g., LoRa \cite{CentenaroZorzi2016,Bor2016}. 

For the wireless channel, we consider a power-law path loss model with exponent $\alpha$. 
 All channels experience independent Rayleigh fading with unit power, i.e., the channel power gain between the device and a BS is $h\sim\exp(1)$, whereas the channel between an interfering device and a BS is $f\sim\exp(1)$. 

\subsection{Transmission access cases}
Since the transmissions are sent over extremely narrow channels, it is difficult to send a signal at a specific channel and carrier frequency. In fact, the frequency offset in a low-cost oscillator could be higher than the signal bandwidth, making it impractical to assume a perfectly channelized system for UNB networks. For this reason, UNB networks are commonly assumed to have an unslotted frequency access \cite{Do2014}. In this paper, we generalize the analysis, considering access to be either slotted or unslotted in time and/or frequency. 

\begin{figure}[t!]
	\center
	\includegraphics[width=1.95in]{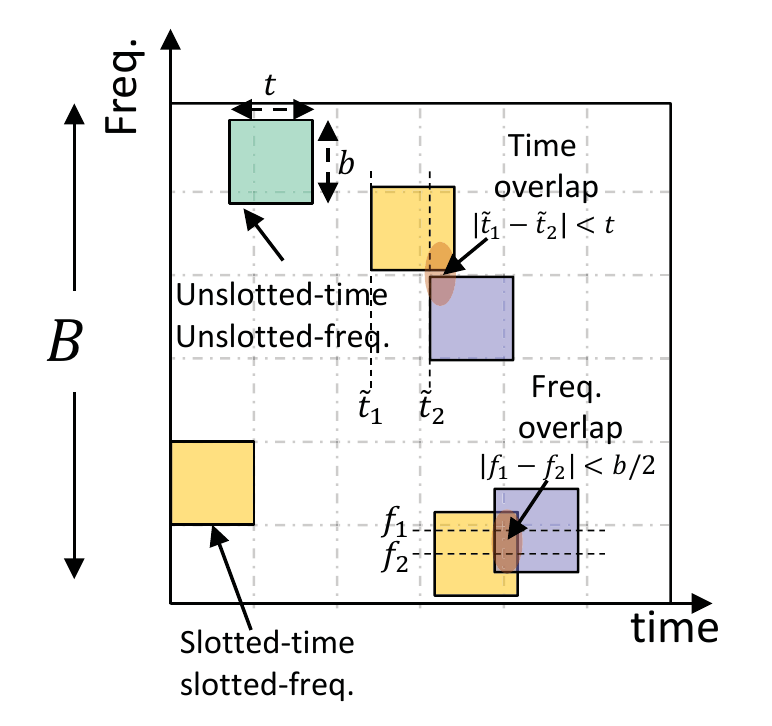}
	\caption{Access can be slotted or unslotted in time and/or frequency.}
	\label{fig:accessScenarios}
	\vspace{-0.15in}
\end{figure}

In the slotted case, a device interferes with another if both devices are transmitting over the same slot, which occurs with probability $t/T$ and $b/B$ in time and frequency, respectively. In the unslotted time (or frequency) case, interference between two devices occur if their signals have any overlap in time (or frequency). For instance, if $\tilde t_1$ and $\tilde t_2$ are the start transmission times of device one and two, then an interference occurs if $|\tilde t_1-\tilde t_2|<t$, and thus two devices interfere with probability $2 t/T$. Similarly, if  $f_1$ and $f_2$ are the center frequencies of the two devices, then an interference occurs if $|f_1 - f_2|<b/2$, which happens with probability $2b/B$. In other words, an unslotted system in time (or frequency) doubles the probability of a device interfering with another one in comparison with a slotted system in time (or frequency \cite{Mo2016}). In this paper, we use $1\leq \beta_{T}\leq 2$, where $\beta_{T}=1$ and $\beta_{T}=2$ denote a slotted and an unslotted system in time, respectively. Note that $\beta_T\in(1,2)$ can denote different levels of time synchronization or interference overlap tolerance. An identical notation is used for frequency, where we use $\beta_F$. An illustration of the different access scenarios is given in Fig. \ref{fig:accessScenarios}.

\subsection{Repetition schemes}
When $N=1$, the signal is sent once, without any repetition. When $N>1$, we consider a \emph{random repetition} scheme, where each IoT transmission is sent at a randomly picked channel, i.e., the IoT device selects $N$ randomly picked channels for packet transmission. In this case, each transmission could interfere with a different set of interfering IoT devices. The impact of this \emph{interference diversity} will be analyzed by comparing the random repetition scheme with another repetition scheme, where the different transmissions of a device will have the same set of interfering IoT devices, i.e., if two devices pick the same first channel, and transmit with time overlap, then both devices will interfere in each transmission. We refer to this scheme as the \emph{Pseudorandom (PN)} scheme, where we assume there exists a set of predetermined sequences, each determining the $N$ channels to be used such that each sequence has channels orthogonal to all other sequences. The number of sequences, over a given time slot, is equal to the number of channels, and thus the probability that two devices, that transmit with some time overlap, pick the same sequence is the same as the probability of two devices picking the same channel in the random repetition scheme.


\subsection{UL access protocols}
\begin{figure*}[t!]
	\center
	\includegraphics[width=7in]{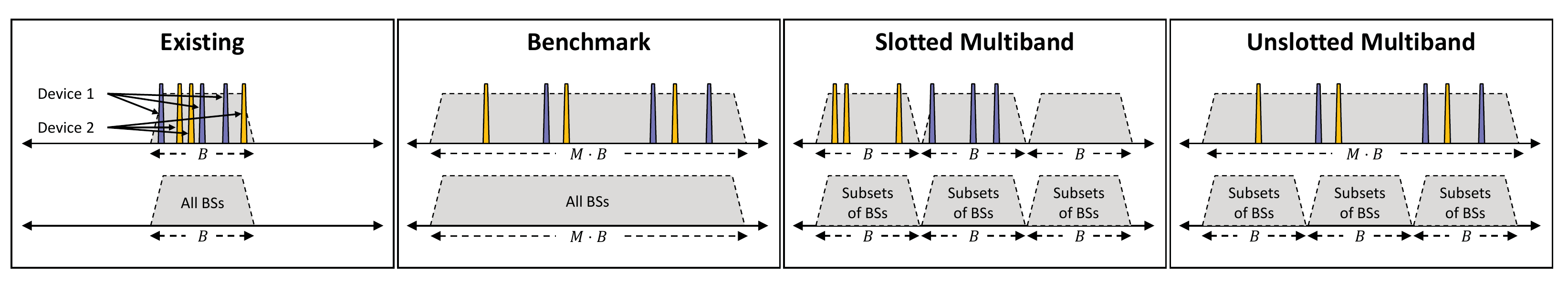}
	\caption{Different access protocols are considered. The top and bottom axes refer to the device side and the BS side, respectively.}
	\label{fig:accessProtocols}
\end{figure*}

Using a wide multiplexing band can help reduce the intra- and inter-network interference. This follows in the former because IoT devices have more channels to pick from, reducing collisions, whereas the latter follows because IoT devices can have a higher probability of using a channel not occupied by an incumbent. However, UNB networks are typically asynchronous in frequency, and thus the UNB receiver must sample the spectrum at a very high resolution to detect IoT signals. For this reason, existing UNB networks assume that devices send their signals within a single narrowband. For example, in a Sigfox system \cite{Sigfox2017}, the BS listens to a spectrum of bandwidth $B=200$KHz, where the power spectral density of the band is obtained using the Fast Fourier transform (FFT), with a very small sampling interval \cite{Artigue2017}. For example, assuming that the device sends a signal of bandwidth $b=100$Hz in this multiplexing band, then using a sampling interval of $b/4$ \cite{Artigue2017} requires the FFT size to be at least $2^{14}$. In other words, using wider multiplexing bands can significantly increase the computational complexity at the receiver. Thus, one question we aim to answer in this paper is whether increasing $B$ to $M\cdot B$ can improve the network performance when each BS is still restricted to sense a single multiplexing band. In particular, we compare between the following access protocols:

\begin{itemize}
	\item \textbf{Existing (or single-band)}: This protocol is similar to existing solutions, e.g., Sigfox, where there is a single multiplexing band, i.e., $M=1$, and all BSs in the network listen to the same band.
	\item \textbf{Benchmark}: This protocol generalizes the Existing one, where there are $M>1$ multiplexing bands, and every BS listens to all of them. While this protocol is impractical for $M\gg1$, it is considered as a benchmark to the best that we can do when receiver complexity is ignored.  	 
	\item \textbf{Slotted Multiband}: In this protocol, we consider $M>1$ multiplexing bands. Each device randomly selects a band and transmits its $N$ messages within the selected band. Similarly, each BS randomly selects a single band to listen to.
	\item \textbf{Unslotted Mulitband}: We consider $M>1$ multiplexing bands. However, each message of the $N$ messages can be sent to a different band. In addition, each BS randomly selects a single band to listen to.  
\end{itemize}
We note that the difference between the slotted and unslotted multiband protocols is that in the former, the same set of BSs will listen to all $N$ messages, whereas in the latter, each message could be received at different subsets of BSs. All these protocols are illustrated in Fig. \ref{fig:accessProtocols}.

\subsection{Performance metrics}
We compare the aforementioned access protocols in terms of \emph{the success probability}, $\mathbb{P}_s$, which is defined as the probability that at least one of the $N$ messages is decoded successfully, where a successful decoding is achieved if the SINR is above a certain threshold $\tau$. Unlike the collision metric, the success probability takes into account the intra- and inter-network interference. Further, the success probability can be interpreted as the complementary cumulative distribution function (CDF) of the maximum SINR among the $N$ messages. This expression is also useful to determine \emph{the transmission capacity} of the network, which is defined as the maximum number of devices that can be supported for a given density of incumbents and success probability constraint $\gamma\in(0,1)$. More formally, the success probability can be written as a function of the density of IoT devices, i.e., $\mathbb{P}_s = F(\lambda_{\IoT})$, and thus, the transmission capacity is defined as $\mathbb{C}(\gamma) = \gamma \cdot F^{-1}(\gamma)$ \cite{Weber2012}.

\section{Performance Analysis of Access Protocols}\label{sec:analysis} 
In this section, we analyze the performance of the access protocols given in Fig. \ref{fig:accessProtocols}. While we focus on asynchronous access with no BS association, we also analyze the performance with nearest BS association to highlight the impact of no BS association on the UNB network performance.

\subsection{Benchmark with no BS association} 
We focus on the benchmark protocol, as the existing one is a special case of the benchmark. The transmission is successful if at least one BS decodes at least one of the $N$ messages. Thus, the success probability is 
\begin{equation*}
\mathbb{P}_s  = 1 - \text{Pr}\{\text{No message is successfully decoded at any BS}\}.
\end{equation*}
Clearly, a BS cannot decode any of the $N$ messages if the message with the maximum SINR is below the decoding threshold $\tau$. More formally, consider a typical device at a distance $x_{j}$ from the $j$-th BS. Then, the SINR of the $i$-th message at this BS can be expressed as
\begin{equation}
\begin{aligned}
\operatorname{SINR}_{i,j} = \frac{h_i x_j ^{-\alpha} }{\hat P_N +\underset{I_{\operatorname{UNB}}}{\underbrace{\sum_{u\in \tilde \Phi_{\IoT}} f_u y_{u,j}^{-\alpha} }} + \underset{I_{\operatorname{INC}}}{\underbrace{\sum_{k\in\tilde \Phi_{\I}} \hat P_{\I} f_k y_{k,j}^{-\alpha}}}},
\end{aligned} 
\end{equation}
where $\hat P_{\I}= \frac{P_{\I}\cdot b/B_{\I}}{P_{\IoT}}$, $\hat P_N= \frac{P_N}{P_{\IoT}}$, $P_N$ is the noise power, $y_{u,j}$ and $y_{k,j}$ are the distances from the $j$-th BS to an interfering UNB device and an interfering incumbent, respectively, and $\tilde \Phi_{\IoT}$ and $\tilde \Phi_{\I}$ are the set of interfering UNB devices and incumbents, respectively. Thus, the $j$-th BS fails to decode the $N$ messages with probability 
\begin{equation}
\label{eq:failProb}
\mathbb{Q}_j = \text{Pr}\left( \operatorname*{max}_{i\in\{1,2,\cdots,N\}} \operatorname{SINR}_{i,j} \leq \tau \right),
\end{equation}
which means that the success probability is given as
\begin{equation}
\label{eq:PsGeneralNoAssociation}
\mathbb{P}_s = 1 - \mathbb{E}_{\Phi_{\B}} \left(\prod_{b\in\Phi_{\B}} \mathbb{Q}_j\right).
\end{equation}
We first consider the PN repetition scheme. In this case, the $N$ messages will have the same set of interfering UNB devices $\tilde \Phi_{\IoT}$, and thus we can simplify (\ref{eq:failProb}) as follows
\begin{equation}
\label{eq:failProbPN}
\begin{aligned}
\mathbb{Q}_j^{\PN}  &= \text{Pr}\left(\operatorname{SINR}_{1,j} \leq \tau, \operatorname{SINR}_{2,j} \leq \tau, \cdots, \operatorname{SINR}_{N,j}\leq \tau \right)\\
&\stackrel{(a)}{=}  \mathbb{E}_{\tilde \Phi_{\IoT},\tilde\Phi_{\I},f_u,f_k}\textstyle \left[\text{Pr}\textstyle\big(h_i \leq \tau x_j^\alpha (\hat P_N +I_{\operatorname{UNB}} + I_{\operatorname{INC}})\big)^N\right]\\
&\stackrel{(b)}{=}  \mathbb{E}_{\tilde \Phi_{\IoT},f_u} \left[\left(1-\mathbb{E}_{\tilde\Phi_{\I},f_k}\Big[e^{-\tau x_j^\alpha (\hat P_N+ I_{\operatorname{UNB}}+I_{\operatorname{INC}})}\Big]\right)^N\right],
\end{aligned}
\end{equation}
where $(a)$ follows as each signal is sent over a different channel with independent fading and the set of interfering IoT devices is the same across $N$ transmissions,\footnote{Channels can be assumed to be independent even if frequency hopping is done over the narrowband $B$ because the duration of each UNB signal is long enough to justify different channels across messages \cite{Sigfox2017}.} and $(b)$ follows from the CDF of $h_i$. Note that an assumption here, that we make throughout the paper, is that each IoT transmission will experience an independent set of interfering incumbents. This assumption simplifies the analysis and is reasonable as we assume incumbent transmitters have shorter transmission durations (recall $B_{\I}\gg b$). For instance, Sigfox transmission duration is $347$ms in the US (or $2$s in Europe), whereas other incumbents, e.g. LoRa, typically have durations of order of few tens of milliseconds for similar packet sizes \cite{SorninHersent2016}.

In the random repetition scheme, each transmission can experience a different and independent set of interfering UNB devices. Thus, we have
\begin{equation}
\label{eq:failProbR}
\begin{aligned}
\mathbb{Q}_j^{\R}  &= \left(\mathbb{E}_{\tilde \Phi_{\IoT},\tilde\Phi_{\I},f_u,f_k} \left[(1-e^{-\tau x_j^\alpha (\hat P_N +I_{\operatorname{UNB}} + I_{\operatorname{INC}})})\right]\right)^N.
\end{aligned}
\end{equation}
To obtain a closed form expression of (\ref{eq:failProbPN}) and (\ref{eq:failProbR}), we need to determine the spatial process of interfering UNB devices, i.e., $\tilde \Phi_{\IoT}$, and interfering incumbent devices, i.e., $\tilde \Phi_{\I}$. Since UNB devices randomly pick time-frequency slots, the set of interfering devices $\tilde \Phi_{\IoT}$ is essentially a thinned $\Phi_{\IoT}$, which is still an HPPP. The density of this thinned process is 
\begin{equation}
\label{eq:interferenceIoTDensity}
\tilde{\lambda}_{\IoT} = N\cdot \beta_{T}\lambda_{\operatorname{T}} \cdot \frac{\beta_{F}b}{M\cdot B}\cdot \lambda_{\IoT}.
\end{equation}
This can be derived as follows. The original density of IoT devices is $\lambda_{\IoT}$, and thus the portion of these devices that would have overlapping transmissions in time is $N \beta_{T} \lambda_{\operatorname{T}}$. However, not all of those devices will have the same frequency as the typical device. Indeed, the probability of a device picking the same frequency as the typical one is $ \frac{\beta_{F}b}{M\cdot B}$, for which (\ref{eq:interferenceIoTDensity}) follows. The same follows for the PN scheme, as the probability of picking the same sequence is $ \frac{\beta_{F}b}{M\cdot B}$.  In a similar manner, the set of interfering incumbents is another HPPP process with density $\tilde{\lambda}_{\I}= \min\{1,\frac{B_{\I}}{M\cdot B}\}\lambda_{\I}$. The next theorem provides the success probability of the benchmark and single-band cases.

\begin{theorem}\label{th:Ps_B}
In an interference-limited network, where $\hat P_N \rightarrow 0$, the success probability of the benchmark protocol under no BS association and PN repetition is given as 
\begin{equation}
\label{eq:Ps_B_PN}
\mathbb{P}_s^{\operatorname{B,PN}} = 1 - \exp\left(\xi \tau^{-\delta}\sum_{k=1}^N \binom{N}{k} \frac{(-1)^k \lambda_{\B}}{k^\delta \tilde{\lambda}_{\IoT}+k\hat P_{\I}^\delta \tilde{\lambda}_{\I}}\right),
\end{equation}
where $\delta=2/\alpha$ and $\xi= \frac{ \sin(\pi\delta)}{\delta\pi}$. If a random repetition scheme is used, then the success probability becomes
\begin{equation}
\label{eq:Ps_B_R}
\mathbb{P}_s^{\operatorname{B,R}} = 1 - \exp\left(-\xi \tau^{-\delta} H_{N}\cdot \frac{\lambda_{\B}}{\tilde{\lambda}_{\IoT}+\hat P_{\I}^\delta\tilde{\lambda}_{\I}}\right),
\end{equation}
where $H_N$ is the harmonic number, i.e., $H_N=-\sum_{k=1}^N \binom{N}{k} (-1)^k k^{-1}$. Further, the performance of the PN scheme  is upper bounded by that of the random scheme. 
\end{theorem}
\noindent \emph{Proof}: See Appendix \ref{app:Ps_B}. $\hfill\blacksquare$

Note that the performance of the existing protocol, i.e., the single-band, can be obtained by setting $M=1$ in (\ref{eq:interferenceIoTDensity}). Additionally, the gain of random repetition over PN repetition follows because in the former, the set of interfering devices differs from one transmission to another, and thus the chances of having a transmission experiencing a lower density of interferers increases. We note that this result follows for all access protocols, as shown in Appendix \ref{app:Ps_B}, and thus we focus on random repetition. 

Using Theorem \ref{th:Ps_B}, we can numerically compute the transmission capacity for the PN repetition scheme, and we can obtain it in closed-form for the random scheme as follows. 
\begin{corollary}
The transmission capacity of the benchmark protocol with a random repetition scheme is 
\begin{equation}
\label{eq:TC_B_R}
\mathbb{C}^{\operatorname{B,R}}(\gamma) = \frac{\gamma M B}{\beta_{T}\beta_{F} b \lambda_{\operatorname{T}}}\left(\frac{\xi \tau^{-\delta} H_N\lambda_{\B}}{N \ln(\frac{1}{1-\gamma})}-\frac{\hat P_{\I}^\delta\min\{1,\frac{B_{\I}}{M B}\}\lambda_{\I}}{N}\right).
\end{equation}	
\end{corollary}
\noindent  We have the following key observations. First, the presence of an interfering network reduces the transmission capacity, as the expression in parentheses decreases with $\lambda_{\I}$. Second, increasing the number of bands has two gains: it can reduce the density of interfering incumbents, i.e., the value in the parentheses is increased, and it scales the UNB transmission capacity, i.e., the term outside the parentheses is also increased. In addition, increasing the number of repetitions has two different effects: it reduces the negative impact of the interfering network, yet it increases the intra-network interference, i.e., the interference from other UNB devices. The former effect follows because sending more transmissions increases diversity, and if $B_{\I}<M\cdot B$, then higher $N$ increases the probability of having at least one message sent in a channel not occupied by an incumbent. The latter effect can be proved by showing that $H_N/N$ is a decreasing function of $N$. In other words, if the interference is dominated by UNB devices, then $N=1$ should provide higher transmission capacity as increasing $N$ increases the UNB interference. However, if the interference is dominated by the incumbent network, then increasing $N$ can be beneficial due to the diversity achieved by sending more messages.   

\subsection{Benchmark with nearest BS association} 
When a device is restricted to connect to the nearest BS, the transmission is considered successful if this particular BS decodes at least one of the $N$ messages. Assuming the nearest BS is $j$, then we have
\begin{equation}
\label{eq:PsGeneralNearestBS}
\begin{aligned}
\mathbb{P}_s  &= 1 - \text{Pr}\{\text{No message is decoded at nearest BS}\}\\
&= 1 - \mathbb{E}_{x} \left(\mathbb{Q}_j\right),
\end{aligned}
\end{equation}
where, unlike (\ref{eq:PsGeneralNoAssociation}), the expectation is with respect to the location of the nearest BS in $\Phi_{\B}$. By evaluating this expression, we get the following theorem. 

\begin{theorem}\label{th:Ps_Bnearest}
	In an interference-limited network, the success probability of the benchmark protocol under nearest BS association and the random repetition is given as
	\begin{equation}
	\label{eq:P_s_U_nearest}
	\bar{\mathbb{P}}_s^{\operatorname{B,R}} = 1-\textstyle \sum_{k=0}^N \binom{N}{k} (-1)^k \left(1+k\xi^{-1} \tau^\delta\frac{\tilde{\lambda}_{\IoT}+\hat P_{\I}^\delta \tilde{\lambda}_{\I}}{\lambda_{\B}}\right)^{-1}.
	\end{equation}
\end{theorem}

\noindent \emph{Proof}: See Appendix \ref{app:Ps_Bnearest}. $\hfill\blacksquare$

The transmission capacity in this case can be computed numerically when $N>1$. When $N=1$, we have the following result.
\begin{corollary}
	The transmission capacity of the benchmark protocol with nearest BS association and $N=1$ is
	\begin{equation}
	\label{eq:TC_B_nearest}
	\bar{\mathbb{C}}^{\operatorname{B}}(\gamma) = \frac{\gamma M\cdot B}{\beta_{T}\beta_{F} b \lambda_{\operatorname{T}}}\left(\frac{\xi \tau^{-\delta} H_N\lambda_{\B}}{(\frac{\gamma}{1-\gamma})}-\hat P_{\I}^\delta\textstyle\min\{1,\frac{B_{\I}}{M\cdot B}\}\lambda_{\I}\right).
	\end{equation}	
\end{corollary}
Comparing (\ref{eq:TC_B_nearest}) with (\ref{eq:TC_B_R}) when $N=1$, and ignoring the presence of an interfering network, we observe that $\mathbb{C}^{\operatorname{B,R}}(\gamma) \propto [\ln(\frac{1}{1-\gamma})]^{-1}$, whereas $\bar{\mathbb{C}}^{\operatorname{B}}(\gamma) \propto (\frac{1}{\gamma}-1)$. Thus, the gap between the transmission capacity under no BS association and the capacity with nearest BS association becomes higher as $\gamma\rightarrow1$, i.e., for stricter success probability constraints, a UNB network with no BS association can help connect significantly more devices compared to a UNB network with each device connected to a single BS.


\subsection{Slotted mutliband with no BS association} 
Let $\mathbb{P}_{s|m}$ denote the success probability given that the typical IoT device picks the $m$-th band for the transmission of $N$ messages. Then, the success probability is 
\begin{equation}
\label{eq:Ps_SM}
\mathbb{P}_{s}^{\operatorname{SM}} = \frac{1}{M}\sum_{m=1}^M  \mathbb{P}_{s|m}.
\end{equation} 
We have the following theorem. 

\begin{theorem}\label{th:Ps_SM}
	In an interference-limited network, the success probability of the slotted multiband protocol with no BS association and random repetition is given as
	\begin{equation}
	\label{eq:SM_R}
	\mathbb{P}_s^{\operatorname{SM,R}} = 1 -  \frac{1}{M}\sum_{m=1}^M  e^{-\xi \tau^{-\delta} H_{N}\cdot \frac{p_m\lambda_{\B}}{\tilde{\lambda}_{\IoT}+\hat P_{\I}^\delta\tilde{\lambda}_{\I}}}.
	\end{equation}
\end{theorem}
\noindent \emph{Proof}: We can compute $ \mathbb{P}_{s|m}$ as follows. The set of BSs listening to the $m$-th band is an HPPP process with density $p_m\lambda_{\B}$. Furthermore, the set of interfering IoT devices is the same as $\tilde \Phi_{\IoT}$. To show this, recall that the density of IoT devices with a time overlap is $N\beta_{T}\lambda_T\lambda_{\IoT}$. Among these devices, only $1/M$ of them, on average, will select the $m$-th band. Among those that select the $m$-th band, only $\beta_{F} b/B$ will select, on average, the same channel as the typical IoT device. Finally, the set of interfering incumbents remains the same, as we assume the incumbent can use any part of the spectrum. To this end, we can use the success probability of the benchmark protocol to compute $\mathbb{P}_{s|m}$, where we replace $\lambda_{\B}$ in (\ref{eq:Ps_B_R}) with $p_m \lambda_{\B}$. $\hfill\blacksquare$

\noindent \emph{Remark}: We note that $\mathbb{P}_s^{\operatorname{SM,R}}$ is maximized when $p_m=1/M$. This can be proved using the inequality $\frac{1}{M}\sum_{m=1}^M  \exp(p_m z)\geq \exp(\frac{z}{M} \sum_{m=1}^M p_m)=\exp(z/M)$, which holds with equality when $p_m=1/M$. This is intuitive as devices select a band with equal probability, and thus BSs must follow the same selection procedure to maximize the success performance. An interesting observation is in the absence of an incumbent network, the slotted multiband has the same performance as the single-band protocol when $p_m=1/M$. This shows that while increasing the number of bands decreases the density of interfering devices, the density of BSs listening to the same band also decreases with the same rate, i.e., there is no gain of using more than one band in this case. However, in the presence of the incumbent network, increasing the number of bands can have different impacts, depending on the bandwidth of the interfering network. We can illustrate this in terms of the transmission capacity, which is given in the following corollary. 
\begin{corollary}
	The transmission capacity of the slotted multiband protocol with $p_m=1/M$ and a random repetition scheme is 
	\begin{equation}
	\label{eq:TC_SM_R}
	\mathbb{C}^{\operatorname{SM,R}}(\gamma) = \textstyle\frac{\gamma  B}{\beta_{T}\beta_{F} b \lambda_{\operatorname{T}}}\left(\frac{\xi \tau^{-\delta} H_N\lambda_{\B}}{N\ln(\frac{1}{1-\gamma})}-\frac{\hat P_{\I}^\delta M\min\{1,\frac{B_{\I}}{M\cdot B}\}\lambda_{\I}}{N}\right).
	\end{equation}	
\end{corollary}
\noindent Let $\mathbb{C}^{\operatorname{SB,R}}(\gamma)$ denote the transmission capacity of the single-band, which can be obtained by substituting $M=1$ in (\ref{eq:TC_B_R}). Then, we have the following remarks. 
\begin{itemize}
	\item If $B_I<B$, and hence $B_I<M\cdot B$, then $\mathbb{C}^{\operatorname{SM,R}}(\gamma)=\mathbb{C}^{\operatorname{SB,R}}(\gamma)$. This follows because in both protocols, there will be channels with no interfering incumbents. While the slotted multiband will have more of such channels, the density of BSs listening to a particular band is reduced.
	\item If $B_I>M\cdot B$, then $\mathbb{C}^{\operatorname{SM,R}}(\gamma)<\mathbb{C}^{\operatorname{SB,R}}(\gamma)$. This follows because both protocols will have the same density of interfering incumbents. However, in slotted multiband, fewer BSs listen to the same band compared to the single-band protocol. 
	\item If $B_I>B$ and $B_I<M\cdot B$, then interestingly $\mathbb{C}^{\operatorname{SM,R}}(\gamma)<\mathbb{C}^{\operatorname{SB,R}}(\gamma)$.  Note that when $B_I>B$, then irrespective of the channels used by the device in the single-band access, there will always be interference from the incumbent network. For the slotted multiband, the density of BSs listening to a particular band is reduced, and as $B_I$ increases, fewer channels will be incumbent-free compared to the case when $B_I<B$. 
\end{itemize}
To summarize, in the slotted multiband protocol, there is a cost due to randomly assigning each BS a single band to listen to, and thus increasing the number of bands is not beneficial, particularly when the bandwidth of the incumbent signal overlaps with all bands. 

\subsection{Unslotted mutliband with no BS association}
In this access scheme, each message can be sent over a different multiplexing band. Assuming that the typical device sends $n_m$ of the $N$ messages over the $m$-th band, where $\sum_{m=1}^M n_m= N$, then the probability of having at least a single message decoded successfully at any BS with the 
random repetition scheme is given as
\begin{equation}
\mathbb{P}_{s|\{n_m\}_{1}^M}^{\operatorname{R}} = 1 - \prod_{m=1}^M \exp\left(-\xi \tau^{-\delta} H_{n_m}\cdot  \frac{p_m\lambda_{\B}}{\tilde{\lambda}_{\IoT}+\hat P_{\I}^\delta\tilde{\lambda}_{\I}}\right).
\end{equation}
This expressions follows because the $m$-th band has $p_m \lambda_{\B}$ BSs listening to $n_m$ messages. Let $\mathcal{N}=\{n_1,n_2,\cdots,n_M|\sum_{m=1}^M n_m=N, 0\leq n_m\leq N\}$ be the set of all possible combinations of sending $N$ messages over $M$ bands. Then, the success probability of the unslotted multiband can be computed by averaging over $\mathcal{N}$, as given in the following theorem.


\begin{theorem}\label{th:Ps_UM}
	In an interference-limited network, the success probability of the unslotted multiband protocol with no BS association and random repetition is given as
	\begin{equation}
	\label{eq:UM_R}
	\mathbb{P}_s^{\operatorname{UM,R}} = \frac{1}{M^N}\sum_{\{n_m\}\subset\mathcal{N}} \frac{N!}{n_1!n_2!\cdots  n_M!}\mathbb{P}_{s|\{n_m\}_{1}^M}^{\operatorname{R}},
	\end{equation}
	where the sum is over every possible combination in $\mathcal{N}$. 
\end{theorem}
\noindent \emph{Proof}: Observe that the probability of sending $n_m$ messages out of $N$ over the $m$th band follows the multinomial distribution, where $N$ is the number of trials and $M$ is the number of possible outcomes for each one. Since each channel is picked with probability $1/M$, then a specific combination $\{n_m\}_{m=1}^M\subset\mathcal{N}$ occurs with probability $\frac{1}{M^N} \frac{N!}{n_1!n_2!\cdots  n_M!}$, which completes the proof. $\hfill\blacksquare$

The transmission capacity of this protocol can be obtained numerically using the aforementioned expressions. 

\subsection{Multiband access with nearest BS association}
In this section, we consider a special case of the slotted multiband protocol, where the device randomly picks  a single band and sends all of its $N$ messages to the nearest BS listening to that band. The success probability is as follows.

\begin{theorem}\label{th:Ps_SM_nearest}
	In an interference-limited network, the success probability of the slotted multiband protocol under nearest BS association and random repetition is given as
	\begin{equation}
	\label{eq:P_s_SM_nearest}
	\bar{\mathbb{P}}_s^{\operatorname{SM,R}} = 1-\textstyle \frac{1}{M}\textstyle\sum_{m=1}^M \sum_{k=0}^N \binom{N}{k} \frac{(-1)^k}{\left(1+k\xi^{-1} \tau^\delta\frac{\tilde{\lambda}_{\IoT}+\hat P_{\I}^\delta \tilde{\lambda}_{\I}}{p_m\lambda_{\B}}\right)}.
	\end{equation}
\end{theorem}
\noindent \emph{Proof}: The proof follows similar to that in Theorem \ref{th:Ps_Bnearest}, and hence it is omitted. The only difference is that the density of BSs are thinned by $p_m$.  $\hfill\blacksquare$

The transmission capacity can be computed numerically when $N>1$ and in closed form for the following case.
\begin{corollary}
	The transmission capacity of the mutliband access scheme with nearest BS association, random repetition, $N=1$, and $p_m=1/M$ is given as 
	\begin{equation}
	\label{eq:TC_SM_nearest}
	\bar{\mathbb{C}}^{\operatorname{SM,R}}(\gamma) = \textstyle \frac{\gamma M\cdot B}{\beta_{T}\beta_{F} b \lambda_{\operatorname{T}}}\left(\frac{\xi \tau^{-\delta} H_N\frac{\lambda_{\B}}{M}}{(\frac{\gamma}{1-\gamma})}-\hat P_{\I}^\delta\textstyle\min\{1,\frac{B_{\I}}{M\cdot B}\}\lambda_{\I}\right).
	\end{equation}	
\end{corollary}
\noindent The observations regarding how the transmission capacity changes with $M$ is similar to those made about (\ref{eq:TC_SM_R}).


\begin{table}[!t]
	\caption{Main parameters}
	\label{tab:parameters}
	\centering
	\begin{tabular}{|l|l|}
		\hline
		\textbf{Description }  				&  \textbf{Parameters}\\\hline
		IoT signal bandwidth		&$b=600$Hz	\\	
		UNB multiplexing band   &$B=200$KHz\\
		Temporal traffic generation & $\lambda_T=2.8\times 10^{-3}$\\
		Number of transmissions & $N=3$\\
		Number of bands         & $M=5$\\
		IoT Tx power & $P_{\IoT}=14$dBm\\
		IoT density & $\lambda_{\IoT}=50\times 10^3/\lambda_{\B}$\\
		Incumbent bandwidth & $B_I=125$KHz\\
		Incumbent effective density & $\lambda_{\I}= 10^3 \lambda_T/\lambda_{\B}$\\
		Incumbent Tx power & $P_{\I}=14$dBm (over $B_I$)\\
		Noise power & $P_N=-146$dBm (over $b$)\\
		Path loss exponent &$\alpha=3.5$\\\hline
	\end{tabular}
\end{table}

\section{Simulation Results}\label{sec:simulations}
In this section, we validate the theoretical expressions via Monte Carlo simulations, where the performance is averaged out over $10^4$ spatial realizations. Unless otherwise stated, we consider the random repetition scheme and use Sigfox US specifications for the UNB network parameters, which are listed in Table \ref{tab:parameters}. The temporal traffic generation assumes each device sends 6 messages per hour, each with duration $t=347$ms. Note that this is the maximum number of packets per hour a Sigfox system currently supports. We further consider a LoRa-like parameters for the incumbents \cite{Bor2016}, and assume the incumbent devices have a similar temporal traffic generation. Finally, in all figures, we use markers to denote Monte Carlo simulations and lines to denote the theoretical results.

\subsubsection{Impact of synchronization}
\begin{figure}[t!]
	\center
	\includegraphics[width=3.25in]{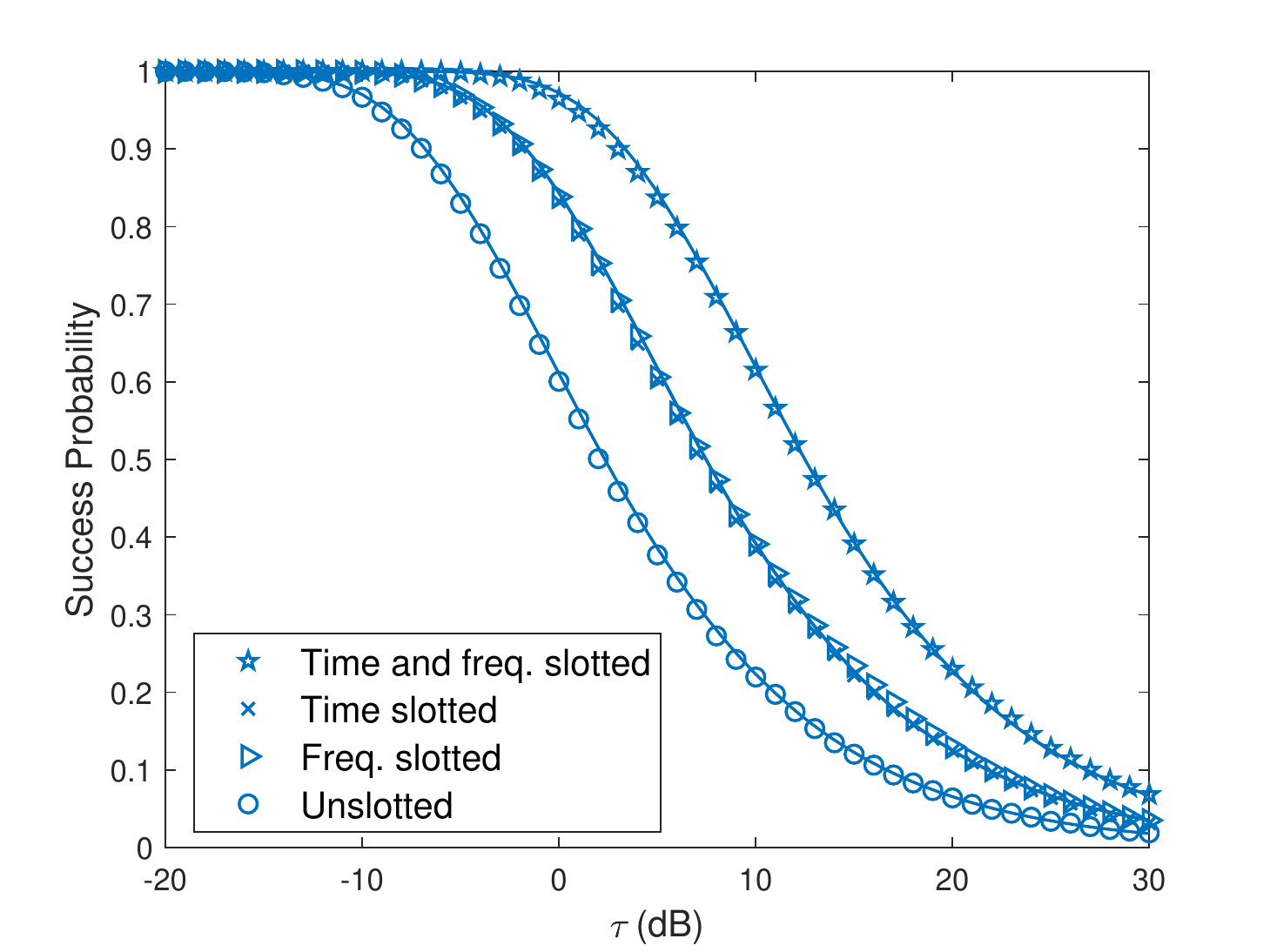}
	\caption{Success probability for the Existing scheme under different access cases.}
	\label{fig:Ps_vs_tau_accessCases}
\end{figure}

We first validate the framework under different transmission access cases, focusing on the Existing protocol (identical trends hold for the different protocols, and hence the results are omitted). Fig. \ref{fig:Ps_vs_tau_accessCases} shows the success probability with variations of the SINR threshold for different access cases, i.e, different levels of time-frequency synchronization. We observe that the theoretical curves match well with simulations. We further make the following observations. First, there is a significant gain achieved when access is slotted in time and frequency, e.g., the median SINR approximately improves by 10dB compared to an unslotted time-frequency system. Second, a time-slotted system has the same performance as a frequency-slotted one. Since frequency synchronization is difficult to achieve for UNB networks, as signals are very narrow, more efforts should be centered around implementing a slotted-time system, particularly because UNB signals span a long duration in time. 

\subsubsection{Success probability comparison}
\begin{figure}[t!]
	\center
	\includegraphics[width=3.25in]{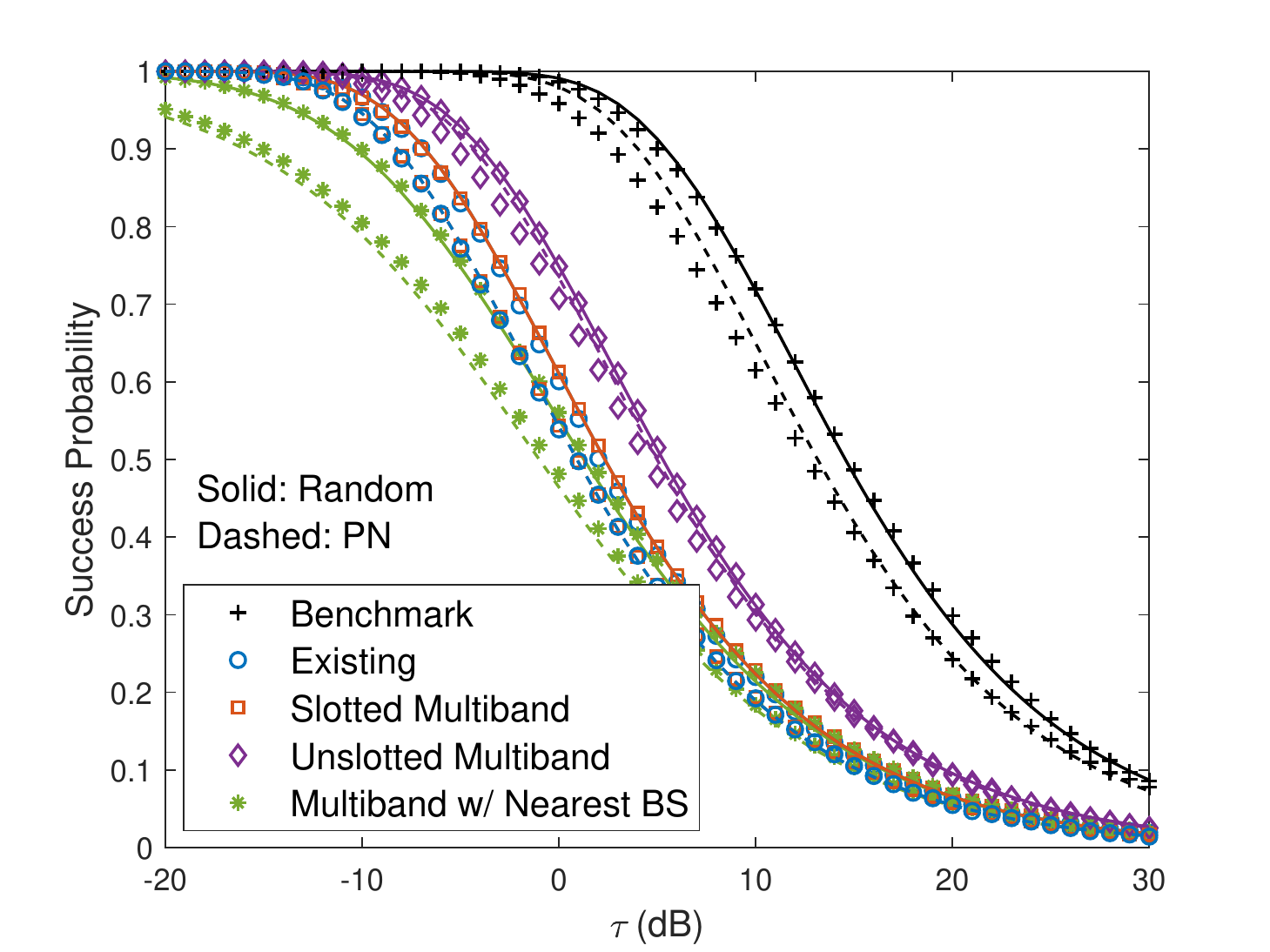}
	\caption{Success probability comparison of different protocols.}
	\label{fig:Ps_vs_tau_protocols}
\end{figure}

We compare the performance of the different protocols in terms of the success probability with variations of the SINR threshold, as shown in Fig. \ref{fig:Ps_vs_tau_protocols}. We have the following remarks. First, the multiband benchmark scheme significantly improves the success probability compared to the Existing protocol, e.g., the median SINR improves by approximately 12dB. However, such protocol is impractical to implement due to the high computational complexity of processing a wider band at a very fine resolution. The unslotted multiband protocol provides a practical compromise, where the median SINR improves by 3dB relative to the single-band scheme, without any additional complexity at the BSs. The unslotted multiband further outperforms the slotted multiband protocol, although the densities of interfering devices and the BSs listening to a given band are the same for both protocols. This follows because in the former protocol, each message could be sent at a different band, and hence each one can be received by a different set of BSs, achieving \emph{a spatial diversity gain} in addition to the diversity achieved by signal repetitions. Further, an access protocol with no BS association improves the cell-edge performance compared to restricting devices to connect to the nearest BS. For instance, the unslotted and slotted multiband schemes improve the cell-edge SINR, i.e., the 5th percentile of the CDF, by 7dB and 4dB, respectively, when compared to a multiband protocol with nearest BS association. 
Finally, interference diversity, resulted from random repetition, is more critical when there are no other diversity sources. Indeed, we observe the random scheme provides 3dB cell-edge improvement over the PN scheme when devices connect to the nearest BS, whereas in the unslotted multiband protocol, with no BS association, the gain is only 1dB because this protocol has another source of diversity, the spatial diversity, as discussed earlier.  

\subsubsection{Impact of the number of bands}
\begin{figure}[t!]
	\center
	\includegraphics[width=3.25in]{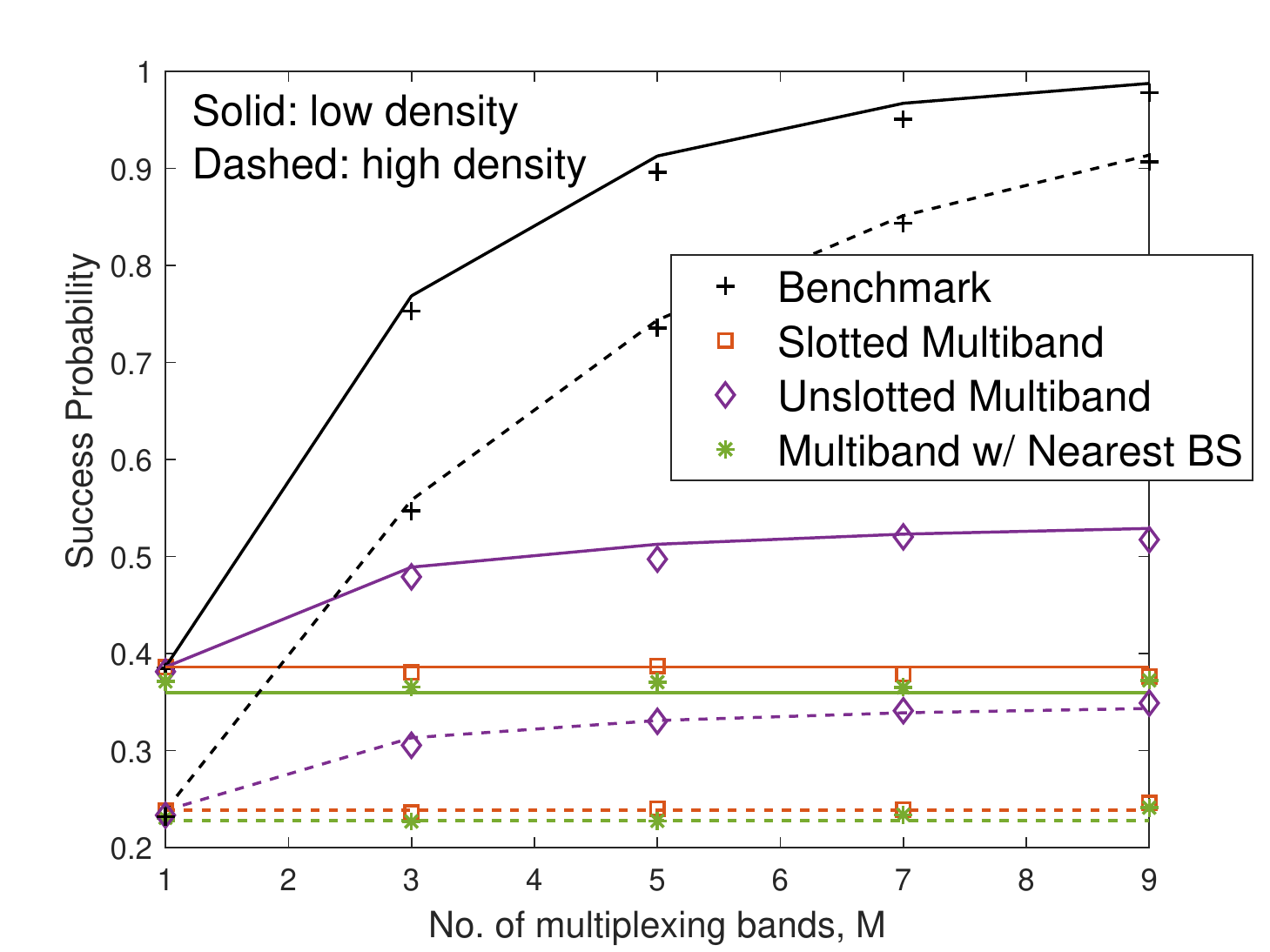}
	\caption{Success probability with variations of $M$.}
	\label{fig:Ps_vs_M}
\end{figure}

We study the success probability performance with variations of the number of multiplexing bands, as shown in Fig. \ref{fig:Ps_vs_M}. Here, we consider two incumbent densities: a low density, where the average number of incumbents is 1000 per BS, i.e., $\lambda_{\I}=10^3 \lambda_T/\lambda_{\B}$, and a high density where the average number of incumbents is the same as the number of IoT devices, i.e., $\lambda_{\I}=\lambda_{\IoT}\lambda_T/\lambda_{\B}$. We observe that the slotted multiband scheme has the same performance as the single-band scheme, and it does not improve with $M$. This follows because increasing $M$ reduces the density of BSs listening to a specific band, which cancels the gain achieved from reducing the density of interferers over a given band. This is not the case with the unslotted mutliband access due to the spatial diversity gain achieved under this protocol, yet the gain saturates with $M$, e.g., increasing the bandwidth of the entire band from 1MHz to 1.8MHz merely improves the success probability by 3\%. Clearly, when BSs are able to listen to all bands, the success probability improves significantly with $M$. Furthermore, the loss due to increasing the density of incumbents by 500x reduces with $M$ for the benchmark and unslotted multiband schemes, with tangible improvements using the former as all BSs listen to all bands. 

\begin{figure}[t!]
	\centering
	\begin{subfigure}[t]{.5\textwidth}
		\centering
		\includegraphics[width=3.25in]{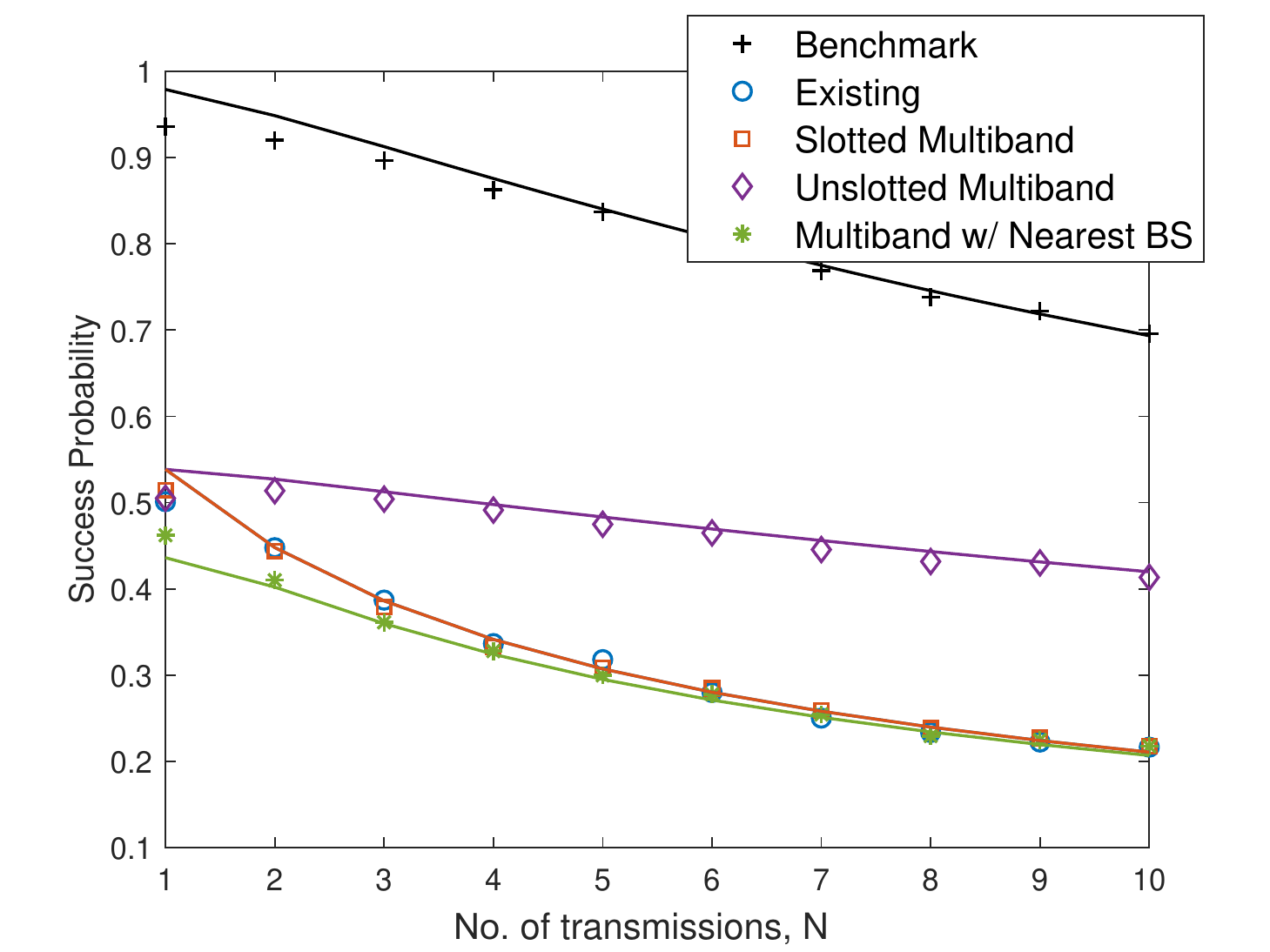}
		\caption{Low density: $\lambda_{\I}=10^3 \lambda_T/\lambda_{\B}$}
		\label{fig:Ps_vs_N_low}
	\end{subfigure}\\
	\begin{subfigure}[t]{.5\textwidth}
		\centering
		\includegraphics[width=3.25in]{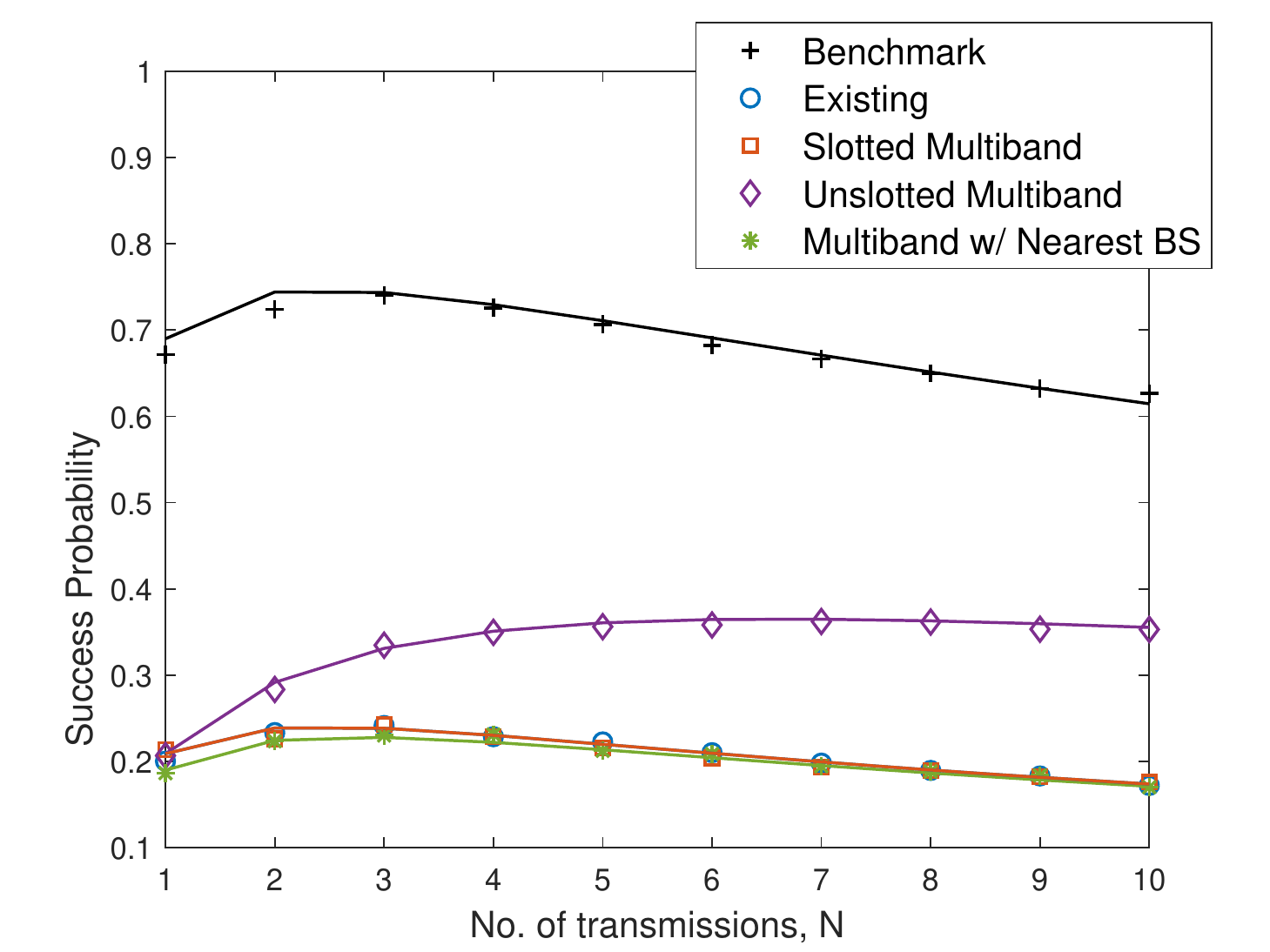}
		\caption{High density: $\lambda_{\I}=\lambda_{\IoT} \lambda_T/\lambda_{\B}$}
		\label{fig:Ps_vs_N_high}
	\end{subfigure}
	\caption{Success probability with variations of $N$.} 
	\label{fig:Ps_vs_N}
\end{figure}

\subsubsection{Impact of the number of repetitions}
We study the success probability performance for different number of transmissions. In Fig. \ref{fig:Ps_vs_N_low}, we show the success probability of the different access protocols with variations of $N$ for low incumbent density. It is observed that when the number of IoT devices is very high relative to the number of channels and the number of incumbent interferers, then increasing $N$ degrades the performance. This follows because the network is dominated by the intra-network interference, and thus to maximize the success probability, each IoT packet should be sent once since increasing $N$ makes the IoT interference higher. However, when the incumbent density is high, the impact of $N$ can change, as shown in Fig. \ref{fig:Ps_vs_N_high}. In this case, the interference becomes dominated by the incumbents sharing the same spectrum, and thus increasing $N$ can provide diversity gain, improving the performance, yet sending too many repetitions eventually degrades the performance. We also observe that the unslotted multiband is more robust than the slotted multiband when the density of incumbents is  high, and it can tolerate higher intra-network interference thanks to the spatial diversity achieved under this protocol.

\subsubsection{Transmission capacity comparison}
We compare the different single-band and practical multiband protocols in terms of the transmission capacity, i.e., how many IoT devices can be supported by a network for a given success probability constraint, where $\tau_{\operatorname{dB}}=5$dB. We assume the network is deployed over an area of $25\times25$km$^2$, and the average number of BSs in that area is $25$. It is shown that UNB networks can provide coverage for a very large number of IoT devices. We observe that the unslotted multiband achieves the highest transmission capacity, where the benefit of spatial diversity is clearly shown when compared to the slotted multiband access. The latter, in fact, has an identical performance to the existing single-band protocol, showing that using $M=5$ improves the transmission capacity by roughly $50\%$ compared to using the existing single-band protocol only when the spatial diversity is exploited. Finally, it is shown that the capacity of unslotted multiband is 2x and 4x the capacity of multiband with nearest BS association for 0.8 and 0.98 success probability constraints, respectively. Thus, no BS association brings significant capacity gains, particularly for high success probability constraints. 

\begin{figure}[t!]
	\center
	\includegraphics[width=3.25in]{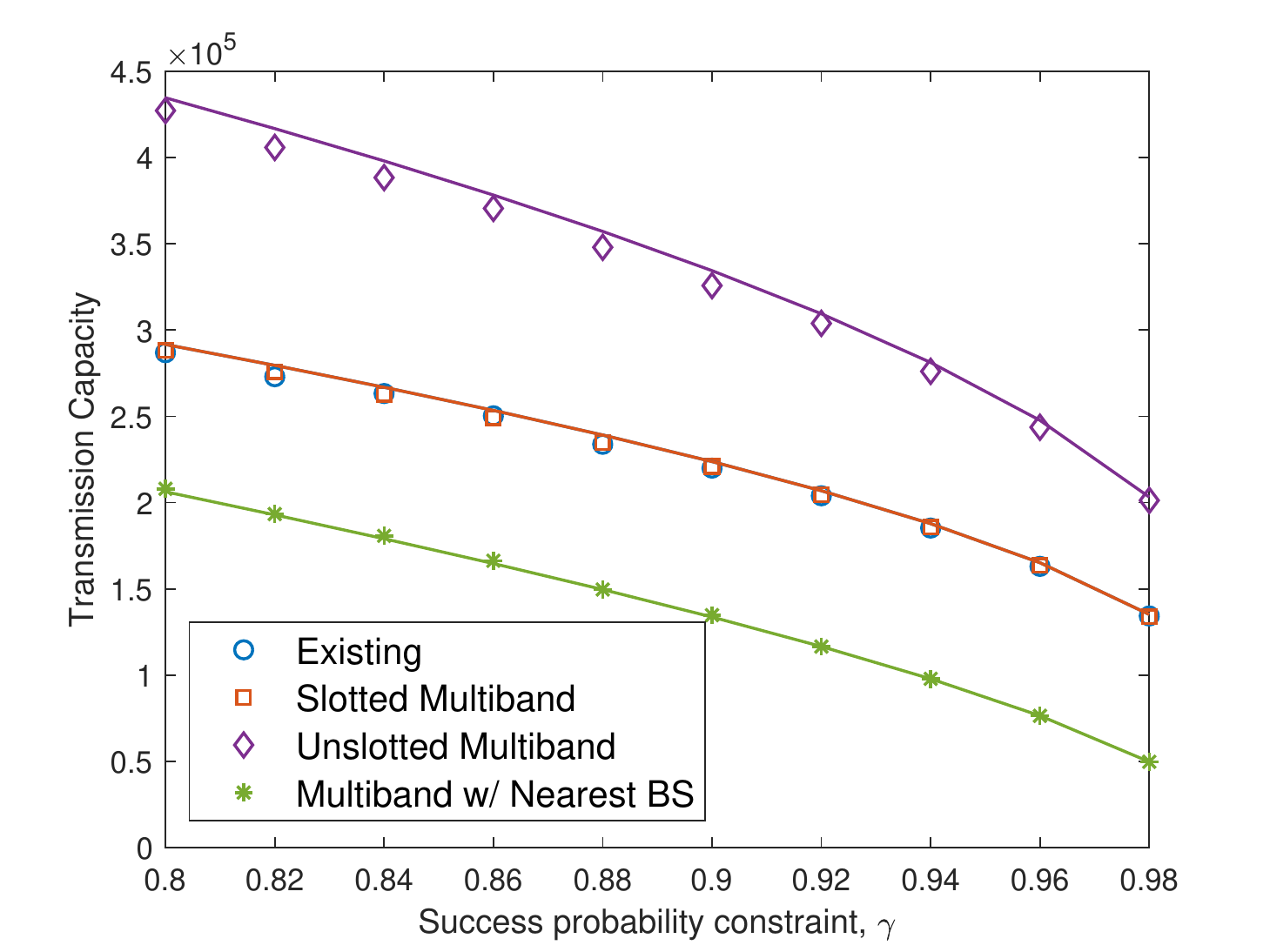}
	\caption{Transmission capacity for a given success probability constraint.}
	\label{fig:TC_vs_gamma}
	\vspace{-.15in}
\end{figure}
\section{Concluding Remarks}\label{sec:conclusion}
In this paper, an analytical framework has been developed to model and analyze UNB communications, an emerging paradigm that relies on ultra-narrowband signals to tackle the intra-network sharing among IoT devices and inter-network sharing for robust coexistence with other incumbent networks. Several access protocols are studied and compared in terms of the success probability and transmission capacity, where closed-form expressions are given for these metrics to identify the key parameters that affect the UNB network performance.

The analysis has shown that the number of signal repetitions and the number of bands play a key role for the success probability. Specifically, sending many repetitions helps improve robustness to the inter-network interference, yet it increases the intra-network interference. Indeed, it is shown that when the network is dominated by interference from IoT devices, a single transmission maximizes the success probability, whereas higher signal repetitions are needed when the density of incumbents increases. In addition, using multiple bands reduces IoT collisions and the interference with incumbents. However, to fully exploit these gains, all BSs are required to listen to all bands. When each BS is restricted to listen to one of the bands, then the diversity achieved by sending several repetitions is not sufficient as the density of UNB BSs listening to a given band is reduced. To mitigate this, each packet transmission should be sent at a different band to exploit a spatial diversity gain, improving the success probability and transmission capacity over single-band protocols. 

Other useful design guidelines have been also gleaned from the analysis. For instance, the interference diversity achieved by random repetition can provide tangible gains over the PN repetition scheme when no other diversity gains are provided by the protocol. In addition, the absence of device-BS association is beneficial for UNB networks as it improves the cell-edge SINR and helps increase network transmission capacity, particularly for high success probability constraints.

\appendix{
\subsection{Proof of the Theorem \ref{th:Ps_B}}\label{app:Ps_B}
\subsubsection{PN repetition}
Consider the PN scheme. We can simplify (\ref{eq:failProbPN}) by taking the average over the incumbent interferes. Specifically, let $s=\tau x_j^\alpha \hat{P}_{\I}^\delta$, then we have
\begin{equation}
\begin{aligned}
\label{eq:INCPGFL}
\mathbb{E}_{\tilde\Phi_{\I},f_k}[e^{-\tau x_j^\alpha I_{\operatorname{INC}}}] &= \mathbb{E}_{\tilde\Phi_{\I},f_k}\left[\exp\left(-s\textstyle \sum_{k\in\tilde\Phi_{\I}}  f_k y_{k,j}^{-\alpha}\right)\right]\\
&\stackrel{(a)}{=}  \exp\left(- 2\pi \tilde{\lambda}_{\I}\textstyle \int_{0}^{\infty}  \big(y-\frac{y}{1+s y^{-\alpha}}\big)dy\right)\\
&= \exp\left(-\pi \xi^{-1} (\tau\hat P_{\I})^\delta \tilde{\lambda}_{\I}x^2 \right),
\end{aligned}
\end{equation}
where $(a)$ follows using the probability generating functional (PGFL) of the interfering incumbent HPPP and the characteristic function (CF) of $f\sim\exp(1)$. We note that the integral limit starts at zero because the interfering incumbent can be arbitrarily close to the BS. Plugging the above expression into (\ref{eq:failProbPN}) and using the binomial theorem, we get
\begin{equation}
\begin{array}{ll}
\label{eq:Q_b_PN}
\mathbb{Q}_j^{\PN}\\
=\mathbb{E}\left[ \sum_{k=0}^N \binom{N}{k} (-1)^k e^{-k(\pi \xi^{-1} (\tau\hat P_{\I})^\delta \tilde{\lambda}_{\I}x^2+\tau x_j^\alpha (\hat P_N + I_{\operatorname{UNB}})) }\right]\\
\stackrel{(a)}{=}\sum_{k=0}^N \binom{N}{k} (-1)^k e^{-\pi \xi^{-1} \tau^\delta (k^\delta \tilde{\lambda}_{\IoT}+k\hat P_{\I}^\delta \tilde{\lambda}_{\I})x^2-k\tau x_j^\alpha \hat P_N} ,
\end{array}
\end{equation}
where $(a)$ follows by taking the expectation with respect to $\tilde \Phi_{\IoT}$ and $f_u$ and using the PGFL of the UNB interfering HPPP, similar to the steps followed in (\ref{eq:INCPGFL}). To compute (\ref{eq:PsGeneralNoAssociation}), we need to take the expectation of (\ref{eq:Q_b_PN}) with respect to the set of all BSs in the network, i.e., $\Phi_{\B}$, as devices do not connect to a single one. Since this set is an HPPP with density $\lambda_{\B}$, we can use the PGFL of the HPPP as follows
\begin{equation}
\label{eq:EQ_b_PN}
\begin{array}{ll}
\mathbb{E}_{\Phi_{\B}}\textstyle\left[\prod_{b\in\Phi_{\B}} \mathbb{Q}_j^{\PN}\right]\\
= \exp \left(-2\pi \lambda_{\B}\int_{0}^\infty x(1-\mathbb{Q}_j^{\PN})dx\right)\\
=\exp\bigg(2\pi \lambda_{\B} \sum_{k=1}^N \bigg\{\binom{N}{k} (-1)^k  \\
\times\int_{0}^\infty x e^{-\pi \xi^{-1} \tau^\delta (k^\delta \tilde{\lambda}_{\IoT}+k\hat P_{\I}^\delta \tilde{\lambda}_{\I})x^2-k\tau x^\alpha \hat P_N} dx\bigg\}\bigg).
\end{array}
\end{equation}
Pluggin (\ref{eq:EQ_b_PN}) in (\ref{eq:PsGeneralNoAssociation}), we get the exact success performance. We can further simplify the expression assuming an interference limited network, i.e., $\hat P_N\rightarrow0$. In this case, we have
\begin{equation}
\label{eq:integralNoNoise}
\begin{aligned}
\int_{0}^\infty x e^{-\pi\xi^{-1} \tau^\delta (k^\delta \tilde{\lambda}_{\IoT}+k\hat P_{\I}^\delta \tilde{\lambda}_{\I})x^2} dx = \frac{ \xi\tau^{-\delta}/(2\pi)}{k^\delta \tilde{\lambda}_{\IoT}+k\hat P_{\I}^\delta \tilde{\lambda}_{\I}}.
\end{aligned}
\end{equation}
Plugging (\ref{eq:integralNoNoise}) in (\ref{eq:EQ_b_PN}) and then using (\ref{eq:PsGeneralNoAssociation}), we arrive at (\ref{eq:Ps_B_PN}). 
\subsubsection{Random repetition}
The difference in this scheme is that we take the expectation first with respect to all random variables, and then apply the binomial theorem. Thus, we have
\begin{equation}
\begin{aligned}
\label{eq:Q_b_R}
\mathbb{Q}_j^{\R} 
&=\left(1-e^{-\pi \xi^{-1} \tau^\delta (\tilde{\lambda}_{\IoT}+\hat P_{\I}^\delta \tilde{\lambda}_{\I})x^2-\tau x_j^\alpha \hat P_N}\right)^N\\
&= \sum_{k=0}^N \binom{N}{k} (-1)^k e^{-k\pi \xi^{-1} \tau^\delta (\tilde{\lambda}_{\IoT}+\hat P_{\I}^\delta \tilde{\lambda}_{\I})x^2-k\tau x_j^\alpha \hat P_N},
\end{aligned}
\end{equation}
and hence
\begin{equation}
\label{eq:EQ_b_R}
\begin{array}{ll}
\mathbb{E}_{\Phi_{\B}}\textstyle\left[\prod_{b\in\Phi_{\B}} \mathbb{Q}_j^{\R}\right]\\
=\exp\bigg(2\pi \lambda_{\B} \sum_{k=1}^N \bigg\{\binom{N}{k} (-1)^k  \\
\times\int_{0}^\infty x e^{-\pi \xi^{-1} k\tau^\delta (\tilde{\lambda}_{\IoT}+\hat P_{\I}^\delta \tilde{\lambda}_{\I})x^2-k\tau x^\alpha \hat P_N} dx\bigg\}\bigg)\\
\stackrel{(a)}{=} \exp\left(\xi \tau^{-\delta} \frac{\lambda_{\B}}{\tilde{\lambda}_{\IoT}+\hat P_{\I}^\delta\tilde{\lambda}_{\I}}\sum_{k=1}^N \binom{N}{k} \frac{(-1)^k}{k} \right),
\end{array}
\end{equation}
where $(a)$ follows under the assumption of an interference-limited network. Thus, plugging (\ref{eq:EQ_b_R}) in (\ref{eq:PsGeneralNoAssociation}), we arrive at (\ref{eq:Ps_B_R}). 

\subsubsection{Comparison between random and PN schemes} 
Next, we prove that the performance of PN repetition is upper bounded by that of the random repetition, for all access protocols. Indeed,  we can rewrite (\ref{eq:failProbPN}) as $\mathbb{E}[g(z)]$, where $g(z)=z^N$, and hence $(\ref{eq:failProbR})$ is rewritten as $g(\mathbb{E}[z])$. Using Jesnen's inequality, we have $g(\mathbb{E}[z])\leq \mathbb{E}[g(z)]$, i.e., $\mathbb{Q}_j^{\R}\leq \mathbb{Q}_j^{\PN}$, since $g(z)$ is a convex function. Thus the success probability of the random repetition scheme outperforms that of the PN scheme.

\subsection{Proof of the Theorem \ref{th:Ps_Bnearest}}\label{app:Ps_Bnearest}

The distribution of the distance to the nearest BS is $f(x)=2\pi \lambda_{\B} x \exp(-\pi \lambda_{\B} x^2)$ \cite{Haenggi2012}. Using random repetition, we have
\begin{equation}
\label{eq:EQ_b_PNnearest}
\begin{aligned}
\mathbb{P}_s^{\operatorname{B,R}}&= 1-\mathbb{E}_{x}\textstyle\left[ \mathbb{Q}_j^{\R}\right]\\
&= 1- 2\pi \lambda_{\B} \sum_{k=0}^N \bigg\{\binom{N}{k} (-1)^k  \\
&\times \int_{0}^\infty  x e^{-\pi x^2 (\lambda_{\B}+k \xi^{-1} \tau^\delta ( \tilde{\lambda}_{\IoT}+\hat P_{\I}^\delta \tilde{\lambda}_{\I}))}dx\bigg\}\\
&=1-\textstyle \sum_{k=0}^N \binom{N}{k} (-1)^k \left(1+\frac{k\xi^{-1} \tau^\delta ( \tilde{\lambda}_{\IoT}+\hat P_{\I}^\delta \tilde{\lambda}_{\I})}{\lambda_{\B}}\right)^{-1}.
\end{aligned}
\end{equation}
We can follow the same procedure to derive the performance with the PN repetition by replacing $\mathbb{Q}_j^{\R}$ with $\mathbb{Q}_j^{\PN}$ in (\ref{eq:EQ_b_PNnearest}). 

}
\bibliographystyle{IEEEtran}
\bibliography{C:/Users/ghait/Dropbox/References/IEEEabrv,C:/Users/ghait/Dropbox/References/References}

\end{document}